\documentclass[epj]{svjour}
\input epsf

\newcounter{append}
\newcommand{\bc}{\begin{center}}
\newcommand{\ec}{\end{center}}
\newcommand{\be}{\begin{equation}}
\newcommand{\ee}{\end{equation}}
\newcommand{\ba}{\begin{array}}
\newcommand{\ea}{\end{array}}
\newcommand{\beqn}{\begin{eqnarray}}
\newcommand{\eeqn}{\end{eqnarray}}

\begin{document}
\title{Off equilibrium dynamics in the 3d-XY system}

\author{St\'ephane Abriet and Dragi Karevski }
\institute{Laboratoire de Physique des Mat\'eriaux, UMR CNRS No. 7556, Universit\'e Henri
Poincar\'e (Nancy 1), B.P. 239,\\ F-54506 Vand\oe uvre l\`es Nancy cedex, France}


\date{Mai 25, 2004}

\abstract{
We investigate through Monte Carlo simulations the non-equilibrium behaviour of the three-dimensional
XY-model quenched from a high temperature state to its ferromagnetic
and critical phases. The two-times autocorrelation and response functions are determined in the asymptotic (scaling) regime, from which
the nonequilibrium exponents $\lambda$ and critical $\lambda^c$ are extracted.
The form of the scaling function is in agreement with the prediction of local scale-invariance.
The so-called limit fluctuation-dissipation ratio $X_\infty$ is shown to vanish in the ordered phase and to reach a constant value
around $0.43$ for the critical quench.
}

\authorrunning{S. Abriet and D. Karevski }
\titlerunning{Off equilibrium dynamics in 3-d XY system}


\maketitle

\section{Introduction}
After the investigations on aging in glass models,\cite{bouchaud,cug2,cug1} efforts have been concentrated
on the study of non-disordered model systems since it was soon realized that some of the basic aspects of aging are present in  
simple systems too with some caracteristic features.\cite{G&L,cug3} In most studies, special attention is put on pure ferromagnetic models
undergoing a second order phase transition. Typically, one  cools down a sample initially prepared in its high temperature phase
to its ordered phase. As time is running, domains start to grow with a linear size $l\sim t^{1/z}$. In the thermodynamic limit,
equilibrium is never reached which implies an infinite relaxation time.\cite{bray} 
In order to probe aging, which implies a full dependence on the past evolution,
it is useful to compute  two-times correlations:
\be
C(t,t_w)=\langle\sigma(t)\sigma(t_w)\rangle \label{defC}
\ee
and linear responses
\be
R(t,t_w)=\left.\frac{\delta\langle\sigma(t)\rangle}{\delta
h(t_w)}\right|_{h=0} \label{defR}
\ee
where $\sigma$ is the order
parameter, $t$ the observation time and $t_w$ the waiting time
($\le t$).\\
Usually, at a coarse-grained level, the dynamics of such systems is described by a Langevin equation:\cite{bray}
\be
\frac{\partial \phi_i(t)}{\partial t}=-\frac{\delta H[\phi]}{\delta \phi_i}
+\eta_i(t)
\label{langevin}
\ee
where $H$ is the free-energy functional, $\eta_i(t)$ is a thermal gaussian
noise at site $i$ with:
\be
\langle\eta_i(t)\rangle=0 \quad,\quad
\langle\eta_i(t)\eta_j(t')\rangle=2T\delta(t-t')\delta_{i,j}\;.
\label{noise}
\ee
As mentioned above, the breakdown of time translation
invariance implies that the two-time functions do not merely depend on
$t-t_w$ but explicitly on $t$ and $t_w$. 
On the bases of general scaling arguments, one postulates the following scaling
laws for the autocorrelation and autoresponse
functions:
\be
C(t,t_w)\approx t_w^{-b}f_C(t/t_w)
\label{genC}
\ee
and
\be
R(t,t_w)\approx t_w^{-1-a}f_R(t/t_w)\;.
\label{genR}
\ee
In the asymptotic limit,
where $t\gg t_w\gg 1$, $f_C$ and $f_R$ decay algebraicly: 
\be
f_C(x)\sim x^{-\lambda_C/z}, \quad f_R(x)\sim x^{-\lambda_R/z}
\label{asy} \ee 
where $z$ is the dynamic exponent and
$\lambda_{C,R}$ are respectively the
correlation \cite{huse} and response exponents.\cite{Picone}

A consequence of the breakdown of time-translation invariance is
the violation of the fluctuation-dissipation theorem (FDT) which relates for equilibrium systems
the response to the correlation function via:
\be
R(t-t_w)=\frac{1}{T}\frac{\partial
C(t-t_w)}{\partial t_w}\;.
\label{FDT}
\ee
When the system is not at equilibrium $R(t,t_w)$ and $C(t,t_w)$ both depend on $t$ and
$t_w$, and the FDT (\ref{FDT}) does not hold anymore. One introduces a new
function $X(t,t_w)$ defined by:\cite{cug3}
\be
R(t,t_w)=\frac{X(t,t_w)}{T}\frac{\partial C(t,t_w)}{\partial
t_w}\;, 
\label{vFDT}
\ee
which measures the deviation from equilibrium.
In some mean-field theory of glassy systems, where it was first formulated,
the fluctuation-dissipation ratio (FDR) $X(t,t_w)$   depends on time only through a functional dependence on
$C(t,t_w)$,\cite{cug2} that is $X=X(C(t,t_w))$. In this case, integrating (\ref{vFDT}) with respect to
$t_w$, one gets $T\chi=\int_{C(t,t_w)}^1X(C')dC'$, so that the slope of the parametric
plot susceptibility $\chi$ versus the correlation function $C$ will give the FDR.
In the asymptotic regime, for such pure ferromagnetic models, the FDR reaches a constant
value $X_{\infty}$ after an initial quasi-equilibrium regime where $X=1$ (FDT valid).
It was argued that this asymptotic constant is universal and characterises the non-equilibrium process.\cite{godreche1}
When a ferromagnetic  system is quenched toward its ferromagnetic phase, 
the long-time dynamics is governed by the diffusion of domain walls and the coarsening of domains.
This coarsening leads to a vanishing FDR $X_\infty=0$. \cite{cug3,barrat,berthier,cannas}.
The scaling functions defined previously are fullfilling this requirement.
In the case of a critical quench, that is exactly at the critical temperature, several studies on various
systems have given some insights. In that case, the reason for aging is related to the development of the
spatial correlations  over a length scale $\xi$ that grows as
$t^{1/z_c}$ where $z_c$ is the critical dynamic exponent. So the
system is still disordered over a length scale larger than
$\xi(t)$, while it looks critical in regions smaller than
$\xi(t)$. In the thermodynamic limit, the equilibration is
unaccessible since the system is of infinite size so $\xi$
indefinitely increases.
For a fully disordered initial state, one has in the kinetic spherical model
$X_\infty=1-2/d$ for $2<d<4$ and $X_\infty=1/2$ for $d\ge 4$,\cite{godreche1}
in the $1d$ Glauber-Ising model $X_\infty=1/2$.\cite{godreche1,lipp}
For algebraic initial correlations, it was shown on the kinetic spherical model
that there exists a rich kinetic phase diagram, depending on the space-dimension $d$ and
the algebraic decay exponent $\omega$,\cite{Picone} where the limit value
$X_\infty$ is either vanishing or depending on $d$ and $\omega$. In the 1d Glauber-Ising model, it was shown that $X_\infty$ is independent
on $\omega$.\cite{henkschutz}
A systematic field-theoretical approach was developped to calculate the limit FDR\cite{calabrese1} and applied to
several models, ferromagnetic systems at criticality, dilute Ising model, model C dynamics.\cite{calabrese,calabrese2,calabrese3}
Simulations in $2d$ systems gave $X_\infty=0.34$ for the Ising model,\cite{chatelain,berthier2}
$X_\infty=0.41$ for the three states Potts model, $X\infty=0.47$ for the 4-states Potts model.\cite{chatelain04}
In $3d$, $X_\infty=0.40$ was obtained for the Ising model.\cite{godreche1}
Recently, performing a Monte Carlo simulations with Glauber dynamics, we found
from the susceptibility versus  correlation function slope a continuously varying FDR
in the 2-d XY model quenched below the Kosterlitz-Thouless point.\cite{abriet2}
But in this case, one has to use with great care this  slope, since the actual definition of the FDR $X_\infty$
leads to a logarithmicaly vanishing result.
In this paper, we present the results obtained on the 3-d XY model quenched onto criticality and below.

\section{The 3d XY-model}
The XY-model is one of
the most popular theoretical models. As it is well-known, in the two-dimensional case
there is no finite magnetization at non-zero temperature, but nevertheless there is
indeed a topological phase transition occuring at finite temperature due to vortex pairing.\cite{berezin,koster}
In three dimensions, the scenario is more conventional since the additional dimension permits the existence of an
ordered phase at finite temperature. Thus, the 3d-system presents a second-order phase transition which is probably
related to the density of vortex-strings \cite{Kohring,Williams}.
The hamiltonian reads:
\be
{\cal H}=-\sum_{\langle i,j\rangle}{\vec{S}_i\cdot\vec{S}_j}
\label{H}
\ee
where $i$, $j$ are nearest neighbour sites on a cubic lattice and the $\vec{S}_i$
are two-dimensional classical spins whose length is set to unity.
From high-temperature series analyses, the predicted critical
inverse temperature is \cite{calTc}: ${\beta_c=1/T_c=0.4539\pm 0.0013}$.
In our simulations, we use the value ${\beta_c=1/T_c=0.4542}$
obtained from a Monte Carlo study \cite{gottlob}. The critical
exponents have been estimated by renormalisation group
techniques and $\epsilon$-expansions.\cite{zinn,fisher} In
our study we take the values of the critical exponents obtained in ref.\cite{torok} from a numerical
approach:
\be
\beta\simeq 0.349\qquad
\nu\simeq 0.672\;. \label{exp}
\ee
Simulations on the three-dimensional XY model with periodic boundary conditions have revealed
that the critical dynamics exponent $z_c$ is close to 2 \cite{minnhagen,mondello}.
For a quench below the critical temperature, that is in the ordered phase, one has $z=2$.\cite{bray}

\section{Numerics}
We simulate the dynamics through a single spin-flip algorithm of Glauber like form where the transition rates $p$
related to the single flip $\{\theta_i\}\rightarrow\{\theta_i'\}$,
where the $\theta$s are the angular variables of the $XY$ model,  reads:
\be
p(\{\theta_i\}\rightarrow\{\theta_i'\})=\frac{\exp(-\beta{\cal H}[\theta_i'])}
{\exp(-\beta{\cal H}[\theta_i])+\exp(-\beta{\cal H}[\theta_i'])}\;
\label{tr}
\ee
with $\beta$ the inverse temperature. One may notice here that the Metropolis algorithm,
with transition rate $p=\min[1,\exp(-\beta \Delta E)]$, 
leads to the same time evolution of the autocorrelation and susceptibilities, as we have checked it.

The two-times autocorrelation function is calculated via
\be
C(t,t_w)=\frac{1}{L^3}\sum_{i}\langle\cos\left[\theta_i(t)-
\theta_i(t_w)\right]\rangle
\label{Cc}
\ee
where the brackets $\langle\,\cdot\,\rangle$ stand for an average over initial
configurations and realisations of the thermal noise.

The two-times linear magnetic susceptibility $\chi(t,t_w)$ can be computed by the application of a random magnetic
field. The amplitude of the field has to be small in order to avoid nonlinear effects.
The ZFC-susceptibility is usually computed utilising a random distributed field via:\cite{barrat}
\be
\chi(t,t_w)=\frac{1}{h^2 L^3}\sum_{i}\langle\overline{h_x\cos\theta_i(t)+
h_y\sin\theta_i(t)}\rangle
\label{ZFC}
\ee
where the line stands for an average over the field realisations.
Indeed, here we concentrate on a different approach which is based on the lines of ref.\cite{chatelain}.
By definition, the linear autoresponse function $R(t,t_w)$ is
\be
R(t,t_w)=\frac{1}{L^3}
\sum_{i}\left.\left({\frac{\delta\langle\cos\theta_i(t)\rangle}
{\delta h_x(t_w)}\!+\!\frac{\delta\langle\sin\theta_i(t)\rangle}
{\delta h_y(t_w)}}\right)\right|_{h=0}\; .
\label{num1R}
\ee
Following ref.\cite{chatelain}, we can reexpress the linear response function
in the form:\cite{abriet2}
\begin{eqnarray}
R(t,t_w)=\beta\langle\cos\theta_i(t)
[\cos\theta_i(t_w+1)-\cos\theta_i^w(t_w+1)]\rangle\nonumber\\
+\beta\langle\sin\theta_i(t)[\sin\theta_i(t_w+1)-\sin\theta_i^w(t_w+1)]\rangle\;,\quad
\label{Rc}
\end{eqnarray}
where
$\cos\theta_i^w$ and $\sin\theta_i^w$ are the components of the local Weiss magnetisation:
\be
S_i^{x,y}=\frac{1}{\beta}\left.\frac{\partial}{\partial h_{x,y}}\ln Z_i\right|_{h=0}\; ,
\label{weiss}
\ee
where $Z_i=\exp\left(-\beta H(\theta_i,h)\right)+\exp\left(-\beta H(\theta'_i,h)\right)$ is the local partition function in the field.
Finally the fluctuation dissipation ratio is calculated via
\be
X(t,t_w)=\frac{TR(t,t_w)}{C(t,t_w+1)-C(t,t_w)}\;.
\label{Xcal}
\ee

\section{Results}
\subsection{Quenches below $T_c$}
We present the results obtained on systems of linear size $L=50$, with periodic boundary conditions, quenched in
 the low-temperature phase from an infinite-temperature initial state. The averages are performed with the use of more than
 $5 000$ samples.
The expected behaviour of the two-times autocorrelation function for coarsening processes is at sufficiently long times \cite{bray}:
\be 
C(t,t_w)=M_{eq}^2(T)
f_C\left(\frac{t}{t_w}\right)\;
\label{CTi}
\ee
where $M_{eq}(T)$ is the equilibrium magnetisation at the temperature $T$.
In order to check this behaviour, we have simulated quenches at two different temperatures,
$T=0.9$ and $T=1.5$, using the Glauber like dynamics. The results are shown respectively on  figure~1 and 2.
\begin{figure}[ht]
\epsfxsize=8cm
\begin{center}
\mbox{\epsfbox{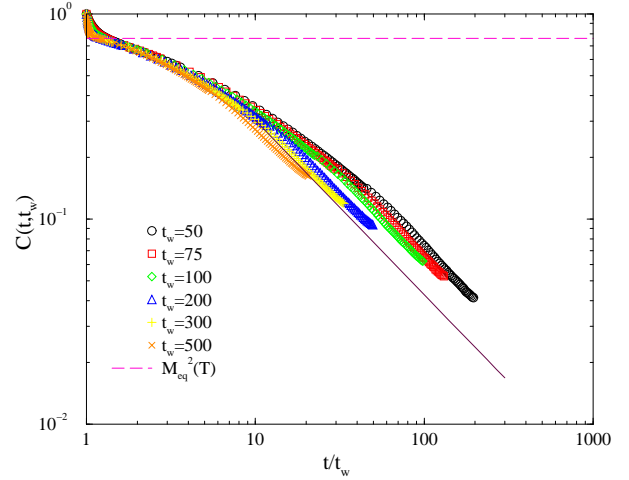}}
\end{center}
\caption{Asymptotic behaviour of the autocorrelation function at
$T=0.9$. The dashed line corresponds to the value $M_{eq}^2(T)$ and
the solid line represents the algebraic decay of $f(t/t_w)$.}
\end{figure}

\begin{figure}[ht]
\epsfxsize=8cm
\begin{center}
\mbox{\epsfbox{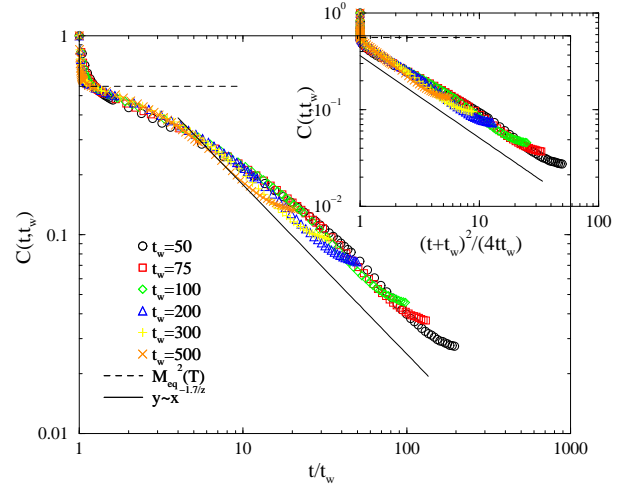}}
\end{center}
\caption{Asymptotic behaviour of the autocorrelation function at
$T=1.5$. The dashed line corresponds to the value $M_{eq}^2(T)$ and
the solid line describes the algebraic decay of $f(t/t_w)$.}
\end{figure}
For small time separations,
we clearly see that the correlations rapidly decay to the expected plateau values
$M_{eq}^2(T)$, here estimated numerically by the use of an independent Wolff algorithm simulation.\cite{wolff,janke}
For large time separations,  the system enters in a scaling region
where the correlation functions are described by a scaling function $f_C$
of the variable $x=t/t_w$. As expected, the aging scaling function finally decays algebraicly
in the asymptotic limit with an exponent $\lambda_C$ roughly about $1.7\pm 0.1$.
However, we clearly see that as the temperature is increased the finite-size effects become stronger.

If one accepts the value $\lambda_C=1.7$, since it is very close to the free field value $\lambda_C^0=d/2=1.5$, one can expect 
a scaling form of the two-times autocorrelation function not too far from the free-field form:\cite{piconehenkel}
\be
C^0(t,t_w)=M_{eq}^2(T)\left(\frac{(x+1)^2}{4x}\right)^{-\lambda_C^0/2}\; ,\qquad x=t/t_w\; .
\ee
Accordingly, we have replotted in the inset of figure~2 the two-times autocorrelation as a function of the scaling variable $y=\frac{(x+1)^2}{4x}$. 
Within this new scaling variable, the power-law form, with $\lambda_C=1.7$, is remarkable.

In the low-temperature phase, the  behaviour of the two-times response function had been conjectured to be
(see Ref. \cite{G&L} for a review):
\be
R(t,t_w)\simeq t_w^{-1-a} f_R\left(\frac{t}{t_w}\right)
\label{RTi}
\ee
where the scaling function $f_R(x)\approx x^{-\lambda_R/z}$ at $x\gg 1$. 
Furthermore, if one assumes that the response function transforms covariantly under local scale-transformations, 
one expects for the scaling function $f_R$ \cite{henkel1,henkel2}
\be
f_R(x)=r_0 x^{1+a-\lambda_R/z}(x-1)^{-1-a}
\label{scR}
\ee
where $r_0$ is a normalization constant.
For a disordered initial state we expect \cite{piconehenkel}
$\lambda_R=\lambda_C=\lambda$, 
with the nonequilibrium  exponent $\lambda$   bounded by\cite{newmann,huse,Yeung}:
$d/2\leq\lambda\leq d$, where $d$ is the euclidian dimension of the system.

We have computed the autoresponse function
 at  temperatures $T=0.9$ and $T=1.5$. On figure~3 and figure~4(left), assuming that $a=1/z=1/2$,\cite{henkelpae} 
the rescaled response function is plotted as a function of $t/t_w$.
\begin{figure}[h]
\epsfxsize=8cm
\begin{center}
\mbox{\epsfbox{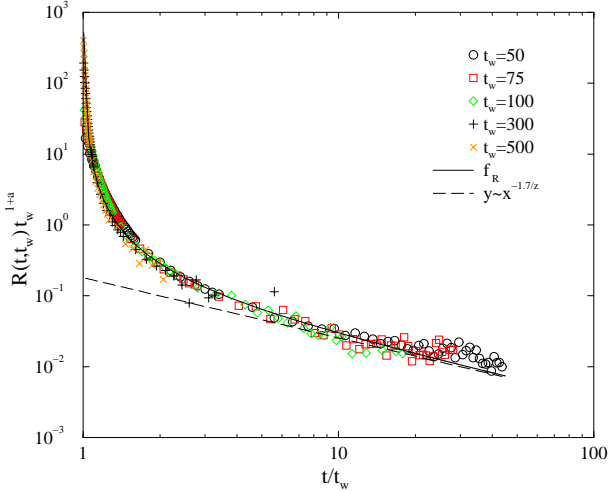}}
\end{center}
\caption{Scaling behaviour of the response function at $T=0.9$ for
different waiting times. The solid line gives the local-scale invariance prediction with $r_0=0.18$. 
The dashed line is a guide to the eyes for the power law behaviour. }
\end{figure}
\begin{figure}[ht]
\epsfxsize=8cm
\begin{center}
\mbox{\epsfbox{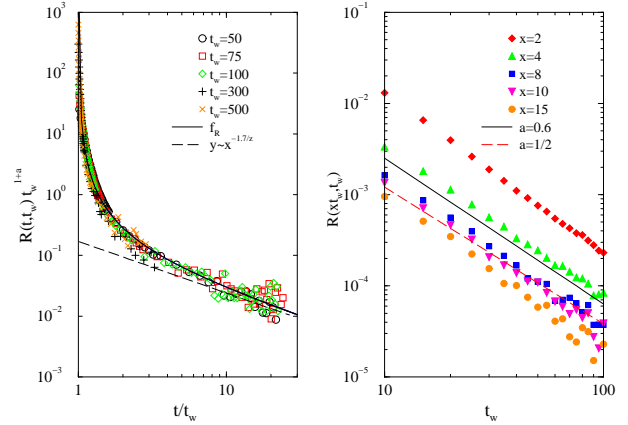}}
\end{center}
\caption{Left: Scaling behaviour of the response function at $T=1.5$ for
different waiting times. The solid line gives the local-scale invariance prediction with $r_0=0.18$. 
The dashed line is a guide to the eyes for the power law behaviour.
Right: Power-law decay of the response function at fixed $x=t/t_w$ ratios and $T=1.5$. 
The results are corroborating the value $a=1/2$.
}
\end{figure}
The collapse of the different waiting time curves is quite good. The value  $a=1/2$ had been already found numerically and analytically in
the 3D Glauber-Ising model \cite{henkel2,berthier} and the 3D kinetic spherical model \cite{Picone,godreche1}.
We have performed an analysis at fixed ratio $t/t_w$ that corroborates the power law prefactor in (\ref{RTi}) with $a$ close to $1/2$
at large $t/t_w$ as seen on figure~4. In fact, numericaly we obtain a value which is slightly larger than $1/2$, around $0.6$ 
but the tendency when increasing $t/t_w$ is toward the decrease of the effective $a$ value.

Due to important thermal fluctuations the algebraic regime $x^{-\lambda_R/z}$ is hardly accessible.  However,
our datas seem to validate the relation $\lambda_C=\lambda_R$ as it can be seen on
figure~3 and figure~4 for $T=0.9$ and $T=1.5$ respectively. Moreover, as it can be seen from figure~3 and 4,
the scaling form (\ref{scR}) predicted from local scale invariance is perfectly reproduced by our datas with $r_0=0.18$.

\begin{figure}[ht]
\epsfxsize=8cm
\begin{center}
\mbox{\epsfbox{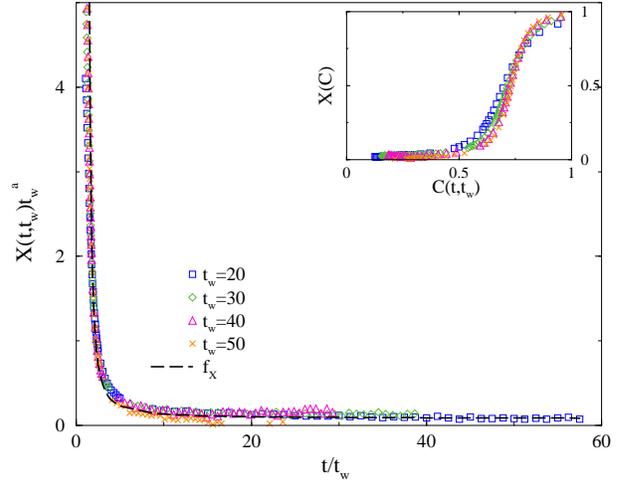}}
\end{center}
\caption{Rescaled Fluctuation-Dissipation ratio $Xt_w^a$ for a quench in the ferromagnetic phase at
 $T=0.9$. The inset is a parametric plot of X versus correlation function. }
\end{figure}
As discussed in the previous section, the asymptotic value of the fluctuation-dissipation ratio
$X_\infty$  is expected to vanish in the low temperature phase.
Assuming that $\lambda_C=\lambda_R$, the
fluctuation-dissipation ratio evaluated from the asymptotic
behaviour of (\ref{CTi}) and (\ref{RTi}) vanishes as:
\be 
X(t,t_w)\simeq
 t_w^{-a}f_X(t/t_w)
 \label{Xasy}
\ee
where $f_X(x)$ is a scaling function related to $f_C(x)$ and $f_R(x)$ with
$\lim_{x\rightarrow \infty}f_X(x)= const.$, a time-independent constant.
On figure~5 we have plotted the rescaled FDR obtained numerically at $T=0.9$, assuming $a=1/2$, for different waiting times
as a function of $t/t_w$. 
 We see a very good collapse of the different waiting time datas. The solid line is the expected scaling form obtained from the  definition of the
 FDR together with the scaling functions $f_C$ and $f_R$  assuming
 the local scale prediction with $r_0=0.18$. Asymptotically, the scaling function $f_X$ 
 reaches a constant  value: 
 \be
\lim_{x\rightarrow \infty} f_X(x)= \frac{Tr_0}{M_{eq}^2(T)\lambda 2^{\lambda-1}}
 \ee
 which is close to $0.08$ at $T=0.9$.
In the inset graph of figure~5, we have represented the FDR $X(t,t_w)$ as a function of the autocorrelation function $C(t,t_w)$ for
different waiting times, where we see the vanishing of the $X$ ratio as the correlation decay.

\subsection{Critical quench}
We focus now on the critical quench starting from a high-temperature
initial state. In the vicinity of the critical point, the
magnetisation at equilibrium behaves as
${|T-T_c|^\beta\sim\xi_{eq}^{-\beta/\nu}}$ and with
$\xi_{eq}\sim t_w^{1/z_c}$, one expects for the two-times autocorrelation function the form:
\be
C(t,t_w)=A^c_C\,t_w^{-2\beta/\nu z_c}
f^c_C\left(\frac{t}{t_w}\right)\;.
\label{Cech}
\ee
The expected scaling for the autoresponse function is given by
\be
R(t,t_w)=A^c_R\,t_w^{-1-2\beta/\nu z_c}
f^c_R\left(\frac{t}{t_w}\right)\; .
\label{Rech}
\ee
The scaling functions $f^c_{C,R}$ are expected to decay algebraicly at
large time separation with the same exponent $\lambda^c_R=\lambda^c_C$, however different from the low-temperature one.

\begin{figure}[ht]
\epsfxsize=8cm
\begin{center}
\mbox{\epsfbox{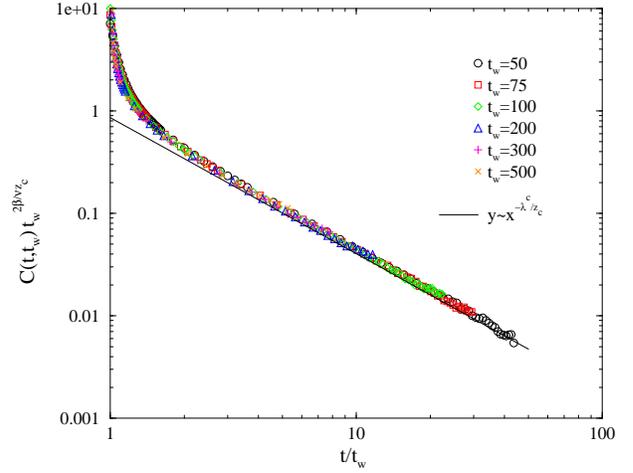}}
\end{center}
\caption{Rescaled autocorrelation functions for a critical quench of the 3D XY-model.
The slope of the solid line stands for $-\lambda^c_C/z_c\simeq -1.34$.}
\end{figure}
The numerical calculations have been performed on periodic cubic lattices of linear size up to 
$L=50$ and the physical quantities have been averaged over
$5000$ noise realisations.
On figure~6 we have plotted the rescaled two-times autocorrelation functions for different waiting times.
The collapse of the rescaled functions is very good, in agreement with
the form (\ref{Cech}). Moreover, at large $t/t_w$ we clearly see a power-law behaviour from which
by an algebraic fit we extracted the exponent $\lambda^c_C/z_c \simeq 1.34$ that fulfills the scaling
bounds $d/2\leq\lambda^c_C\leq d$ (remember that $z_c$ is close to $2$).
\begin{figure}[ht]
\epsfxsize=8cm
\begin{center}
\mbox{\epsfbox{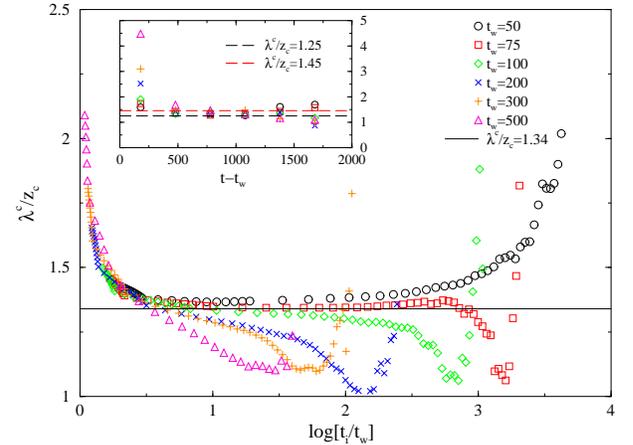}}
\end{center}
\caption{Effective exponent $\lambda_C^c/z_c$ for different waiting times as a function of $t$
obtained from systems of size $L=50$.}
\end{figure}
Analysing more carefully the datas, we obtain a serie of estimates for the exponent
$\lambda^c_C/z_c$ that are represented on figure~7. Our final estimate is
\be
\lambda_C^c/z^c = 1.34 \pm 0.05
\ee
We have also explored the finite-size effects in more details. For that purpose we have studied the time
evolution of cubic systems of linear sizes $L=10, 20, 30$ and $40$. Assuming that
\be
C(t,t_w,L)=b^{-2\beta/\nu}
C(b^{-z_c}t,b^{-z_c}t_w,b^{-1}L)
\ee
with $b=L$ we obtain the scaling form
\be
C(t,t_w,)=L^{-2\beta/\nu}
{\cal F}\left(\frac{t}{L^{z_c}},\frac{t_w}{L^{z_c}}\right)\; .
\label{ech}
\ee
\begin{figure}[ht]
\epsfxsize=8cm
\begin{center}
\mbox{\epsfbox{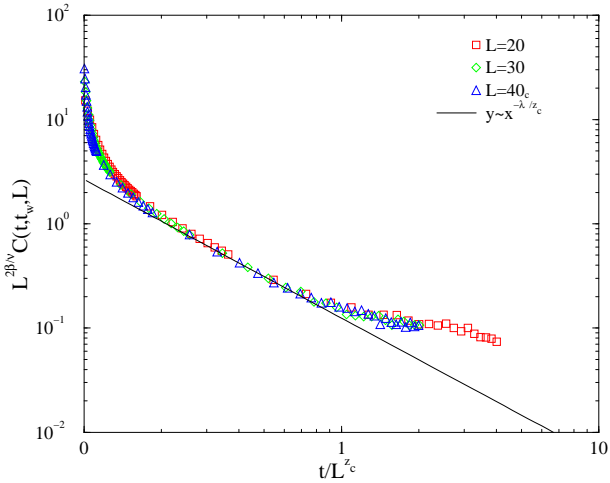}}
\end{center}
\caption{Rescaled autocorrelation functions for different system sizes. All the
curves collapse in the full-aging regime where the scaling function {\cal F} decays
as $(t/L^{z_c})^{\lambda^c_C/z_c}$}
\end{figure}
On figure~8, we have plotted $\cal F$ for different
sizes at fixed ratio $x\equiv t_w/L^{z_c}=0.1$ as a function of $t/L^{z_c}$. Taking into account
(\ref{Cech}) and (\ref{ech}), we expect that the scaling function $\cal F$ has an algebraic decay
 with the exponent $-\lambda^c_C/z_c$ for $t$ not too large. The collapse of the
curves is quite good. For earliest times the system
is still in quasi-equilibrium and for biggest times the growth of correlated
domains is limited by the size of the system. 
The numerical datas seem to validate the scaling assumption (\ref{ech}).

We focus now on the autoresponse functions computed with formula (\ref{Rc}).
The quantities are quite noisy since thermal fluctuations are
of the same order as the amplitude of the response at long times.
From (\ref{genR}), one should have $a=b$ and ${\lambda^c_R=\lambda^c_C=\lambda^c}$
since the initial state is fully disordered. On figure~9, the rescaled autoresponse  is
represented as a function of $t/t_w$ for several waiting times $t_w$. We see that the decay of the autoresponse function is compatible with
the algebraic assumption with ${\lambda^c_R/z_c=\lambda^c/z_c\simeq 1.34}$.
\begin{figure}[h]
\epsfxsize=8cm
\begin{center}
\mbox{\epsfbox{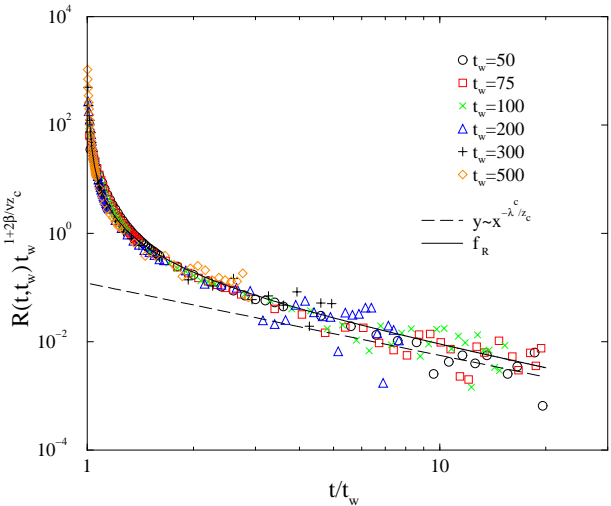}}
\end{center}
\caption{Rescaled autoresponses as a function  of $t/t_w$ for different waiting times and for a critical
quench. The solid line gives the scaling function as predicted by local scale invariance.
The dashed line gives the algebraic decay $(t/t_w)^{-\lambda^c_R/z_c}$ with
$\lambda^c_R/z_c=\lambda^c_C/z_c=1.34$.}
\end{figure}

As stated previously, the out-of equilibrium behaviour is somehow characterised by the FDR $X(t,t_w)$,
calculated numerically via (\ref{Xcal}).
Assuming the scaling forms of the
two-times autocorrelation and response functions, 
the FDR is supposed to be a function of the ratio $t/t_w$ only. 
This behaviour is well reproduced on figure~10, where 
we have plotted the FDR $X(t,t_w)$ as a function of $t_w/t$ calculated on a system of linear size $L=30$, and 
for different waiting times. In the long time limit $t_w/t\rightarrow 0$, the FDR approaches a constant value, around $0.4$, but due to the 
numerical noise, basicaly related to the numerical derivative of $C(t,t_w)$, 
it is difficult to give a more precise value. However, if one supposes an asymptotic linear form
\be 
X(t,t_w)\simeq X_{\infty}+ \delta \frac{t_w}{t}\; ,
\ee
it is possible to extract the value:
\be
X_{\infty}=0.43 \pm 0.04 \; .
\ee
In ref. \cite{calabrese}, the limit FDR was calculated in an $\epsilon$ expansion. The two-loop expansion leads to a
value $X_{\infty}=0.416(8)$ that perfectly fits, within the error bars, our estimate.
\begin{figure}[h]
\epsfxsize=8cm
\begin{center}
\mbox{\epsfbox{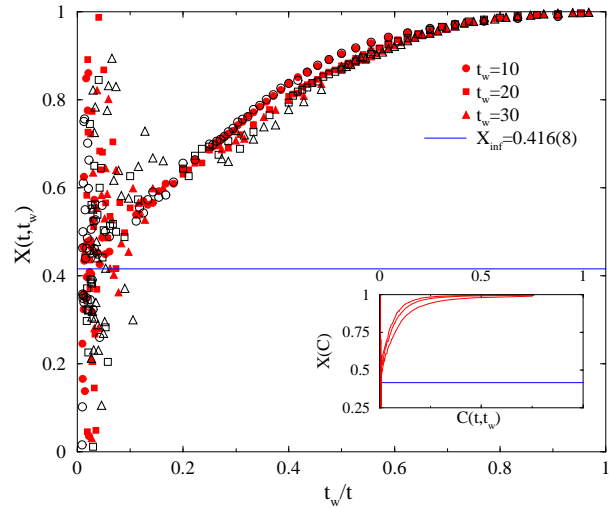}}
\end{center}
\caption{Fluctuation Dissipation ratio at the critical point as a function of $t_w/t$ for $L=20$ (open symbols) 
and $L=30$ (filled symbols). The solid line gives  the analytical result
$X_{\infty}=0.416(8)$ of ref. \cite{calabrese}. The inset shows the dependence of $X$ with the same waiting times
as a function of the correlation $C$.}$\epsilon=4-d$
\end{figure}


\section{Conclusion and outlooks}
We  have studied the relaxation behaviour of the 3D XY-model quenched from
its high temperature state to its ferromagnetic phase and at criticality. Through Monte Carlo simulations,
we have computed the autocorrelation function and its conjugate response function on cubic
lattices of maximal linear size $L=50$.
From our datas, we have confirmed the general scaling scenario at and below criticality. The autocorrelation
non-equilibrium exponent $\lambda_C$ below $T_c$ was found to be around $1.7$. The scaling analysis of the linear autoresponse
function confirmed the equality ${\lambda_R=\lambda_C=\lambda}$, expected for quenches from a fully disordered initial state.
At criticality, we have confirmed that
the correlation and response function behave as:
${C(t,t_w)=t_w^{-b}f_C^c(t/t_w)}$ and
${R(t,t_w)\approx t_w^{-1-a}f_R^c(t/t_w)}$, with $a=b=2\beta/\nu z_c$.
This conclusion has been supported by a finite-size analysis on
systems of linear sizes $L=10, 20, 30, 40$ and $50$.
From the datas on the correlation function we extracted the value
$\lambda^c/z_c\simeq 1.34$ for the nonequilibrium critical exponent and checked its compatibility
with the response function datas. One may notice that this value is
very close to that obtained in the $3d$-Ising model, that is $\lambda^c\simeq 2.8$.\cite{huse}
It is remarkable to notice that the local scale-invariance prediction for the response function is
very nicely fulfilled either at criticality and below. Although at criticality field-theoretical approach\cite{calabrese1} showed a
deviation from the local scale invariance prediction,\cite{calabrese}
it is very difficult to see numerically this very small discrepancy.
In the ordered phase, since the nonequilibrium exponent is very close
to the free-field value $d/2$, we have shown that the autocorrelation function has a scaling
form which is very close to that of the free-field:
$C(t,t_w)=M_{eq}^2y^{-\lambda/z}$ with the scaling variable $y=(x+1)^2/(4x)$ where $x=t/t_w$.
The direct calculation of the FDR $X(t,t_w)$ showed that, as expected, in the low temperature phase it vanishes like
$X_\infty(t,t_w)=t_w^{-a}f_X(t/t_w)$ where $f_X$ is a scaling function simply related to the local scale-invariance
scaling functions $f_C$ and $f_R$.
The same analysis at the critical point gave a limit FDR $X_\infty=0.43$,
which is in agreement,  within the error bars, with the field theoretical value obtained in ref.\cite{calabrese}
To conclude, we have studied the $3d-XY$ model in the context of aging and
showed that the results obtained are consistent with the general picture usually drawn. Moreover, we have given a new verification
of the local scale-invariance predictions, the first for a non-scalar order parameter.

\section*{Acknowledgments}
We wish to thank C. Chatelain and M. Henkel for useful discussions and critical reading of the paper.
P. Calabrese and A. Gambassi are greatfully acknowledged for helpful comments.


\begin{thebibliography}{}
\bibitem{bouchaud} J-P. Bouchaud, L. F. Cugliandolo, J. Kurchan, M. M\'ezard, in {\it Spin Glasses and Random Fields},
edited by A. P. Young (World Scientific, Singapore, 1998);
L. F. Cugliandolo,  Lecture notes, Les Houches, July 2002, cond-mat/0210312
\bibitem{cug2} L. F. Cugliandolo and J. Kurchan J. Phys. A {\bf 27}, 5749 (1994)
\bibitem{cug1} L. F. Cugliandolo, J. Kurchan and L. Peliti Phys. Rev. E {\bf 55}, 3898 (1997)
\bibitem{G&L}  C. Godreche and  J.M. Luck, J. Phys. Cond. Matter {\bf 14}, 1589 (2002);
M. Henkel, Adv. Solid State Phys. {\bf 44} (2004) in press (cond-mat/0404016)
\bibitem{cug3} L. F. Cugliandolo, J. Kurchan and G. Parisi J. Phys. I (France) {\bf 4}, 1641 (1994)
\bibitem{bray} A. J. Bray, Adv. Phys. {\bf 43}, 357 (1994)
\bibitem{huse} D. S. Fisher and D. A. Huse, Phys. Rev. B {\bf 38}, 373 (1988);
D. A. Huse, Phys. Rev. B {\bf 40}, 304 (1989)
\bibitem{barrat} A. Barrat, Phys. Rev. E {\bf 57}, 3629 (1998)
\bibitem{berthier} L. Berthier, J. L. Barrat and J. Kurchan, Eur. Phys. J. {\bf B11}, 635 (1999)
\bibitem{cannas} S. A. Cannas, D. A. Stariolo and F. A. Tamarit, Physica A {\bf 294}, 362 (2001)
\bibitem{godreche1} C. Godreche and J. M. Luck, J. Phys. A {\bf 33}, 9141 (2000)
\bibitem{lipp} E. Lippiello and M. Zannetti, cond-mat/0001103\cite{henkschutz}
\bibitem{henkschutz}  M. Henkel and G. M. Sch\"utz,  J. Phys. A {\bf 37}, 591 (2004)
\bibitem{calabrese1} P. Calabrese and A. Gambassi, Phys.Rev. E {\bf 65}, 066120 (2002)
\bibitem{calabrese} P. Calabrese and A. Gambassi, Phys. Rev. E {\bf 66}, 066101 (2002)
\bibitem{calabrese2} P. Calabrese and A. Gambassi, Phys. Rev. B {\bf 66}, 212407 (2002)
\bibitem{calabrese3}  P. Calabrese and A. Gambassi, Phys.Rev. E {\bf 67}, 036111 (2003)
\bibitem{berthier2} P. Mayer, L. Berthier, J. Garrahan et P. Sollich, Phys. Rev. E {\bf 68}, 016116 (2003)
\bibitem{abriet2} S. Abriet and D. Karevski, Eur. Phys. J. B {\bf 37}, 47 (2004)
\bibitem{berezin} V. L. Berezinskii, Sov. Phys. JETP {\bf 32}, 493 (1971)
\bibitem{koster} J. M. Kosterlitz and D. J. Thouless, J. Phys. C {\bf 6},  1181 (1973);
J. M. Kosterlitz, J. Phys. C {\bf 7},1046 (1974);
J. Villain, J. Physique {\bf 36}, 581 (1975)
\bibitem{Kohring} G. Kohring, R. E. Shrock and P. Wills, Phys. Rev. Lett. {\bf 57}, 1358 (1986)
\bibitem{Williams} G. A. Williams, Phys. Rev. Lett. {\bf 59}, 1926 (1987)
\bibitem{calTc} M. Ferer, M. A. Moore and M. Wortis, Phys. Rev. B {\bf 8}, 5205 (1973)
\bibitem{gottlob} A.P. Gottlob and M. Hasenbusch, Physica A {\bf 201}, 593 (1993)
\bibitem{zinn} J. C. Le Guillou and J. Zinn-Justin, Phys. Rev. B {\bf 21}, 3976 (1980);
R. Guida and J. Zinn-Justin, J. Phys. A {\bf 31}, 8103 (1998)
\bibitem{fisher} K. G. Wilson and M. E. Fisher, Phys. Rev. Lett. {\bf 28}, 240 (1972)
\bibitem{torok} M. Hasenbusch and T. T\"or\"ok, J. Phys. A {\bf 32}, 6361 (1999)
\bibitem{minnhagen} L. M. Jensen, B. J. Kim and P. Minnhagen, Europhys. Lett. {\bf 49}, 644 (2000); 
P. Minnhagen, B. J. Kim and H. Weber, Phys. Rev. Lett. {\bf 87}, 037002 (2001)
\bibitem{mondello} M. Mondello and N. Goldenfeld, Phys. Rev. A {\bf 42}, 5865 (1990);
M. Mondello and N. Goldenfeld, Phys. Rev. A {\bf 45}, 657 (1992)
\bibitem{chatelain} C. Chatelain, J. Phys. A {\bf 36}, 10739 (2003);
F. Ricci-Tersenghi, Phys. Rev. E {\bf 68}, 065104(R) (2003)
\bibitem{chatelain04} C. Chatelain, cond-mat/0404017
\bibitem{Picone} A. Picone and M. Henkel, J. Phys. A {\bf 35}, 5575 (2002)
\bibitem{wolff} U. Wolff, Phys. Rev. Lett. {\bf 62}, 361 (1989)
\bibitem{janke} W. Janke, Phys. Lett. A {\bf 148}, 306 (1990)
\bibitem{piconehenkel} A. Picone and M. Henkel, Nucl. Phys. {\bf B688}, 217 (2004)
\bibitem{henkel1} M. Henkel, Nucl. Phys. {\bf B641}, 405 (2002)
\bibitem{henkel2} M. Henkel, M. Pleimling, C. Godr\`eche and J.-M. Luck,  Phys. Rev. Lett. {\bf 87}, 265701 (2001)
\bibitem{newmann} T. J. Newman and A. J. Bray, J. Phys. A {\bf 23}, 4491 (1990);
J. G. Kissner and A. J. Bray, J. Phys. A {\bf 26}, 1571 (1993)
\bibitem{Yeung} C. Yeung, M. Rao and R. C. Desai, Phys. Rev. E {\bf 53}, 3073 (1996)
\bibitem{henkelpae} M. Henkel, M. Paessens and M. Pleimling, Europhys. Lett. {\bf 62}, 664 (2003)
\end{thebibliography}
\end{document}